\documentclass[aps,prl,twocolumn,groupedaddress,showpacs]{revtex4}
\usepackage{graphicx}

\begin{document}
\title{Static dielectric properties of carbon nanotubes from first
principles} \author{Boris Kozinsky$^{1}$ and Nicola Marzari$^{2}$}
\affiliation{$^{1}$Department of Physics, MIT, Cambridge, MA 02139\\
$^{2}$Department of Materials Science and Engineering, MIT, Cambridge,
MA 02139} \date{\today}
\begin{abstract}
We characterize the response of isolated single- (SWNT) and
multi-wall (MWNT) carbon nanotubes and bundles to static electric
fields using first-principles calculations and density-functional
theory. The longitudinal polarizability of SWNTs scales as the
inverse square of the band gap, while in MWNTs and bundles it is
given by the sum of the polarizabilities of the constituent tubes.
The transverse polarizability of SWNTs is insensitive to band gaps
and chiralities and is proportional to the square of the effective
radius; in MWNTs the outer layers dominate the response. The
transverse response is intermediate between metallic and insulating,
and a simple electrostatic model based on a scale-invariance
relation captures accurately the first-principles results.
Dielectric response of non-chiral SWNTs in both directions remains
linear up to very high values of applied field.
\end{abstract}
\maketitle Carbon nanotubes attract a lot of scientific interest due
to their unique and versatile electronic and mechanical properties,
suitable for a wide range of applications. Nanotubes have
different electronic properties, determined in the zone-folding
scheme by the chiral vector: armchair $(m,m)$ nanotubes are 1D
metals, and zigzag $(m,0)$ nanotubes are semiconductors, with
(almost) vanishing gaps for $m=3n$. Synthesis and separation of
specific nanotubes remains a central challenge. Variations in
chirality and size influence dielectric properties, which in turn
can be exploited for separation; e.g. electric fields have been used
to align nanotubes during PECVD synthesis \cite{dai,zhang} and to
separate different tubes in solutions \cite{solution}. A detailed
physical understanding of dielectric response is also needed to
characterize optical excitations, screening at contacts, plasmons in
nanotube arrays, and the degree of control achievable on endohedral
fillings. While in recent years the response of SWNTs has been
studied with tight-binding \cite{benedict,levitov,rotkin} and
first-principles approaches \cite{nonlinear,scuseria}, MWNTs -- a
more common product of synthesis -- have received much less
attention due to their complexity. We present here a comprehensive
and detailed picture of dielectric screening in SWNTs, MWNTs and
bundles, using a combination of first-principles techniques and
introducing an accurate classical electrostatic model that captures
the unusual response of these materials.

\begin{figure}[htb]
\centering \includegraphics[width=3.0 in]{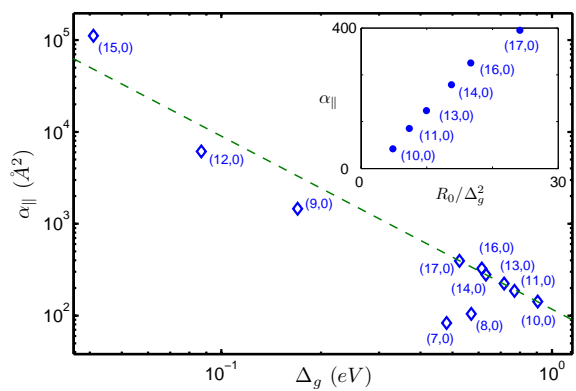}
  \caption {Log scale plot of $\alpha_\parallel$  of
zigzag nanotubes as a function of band gap. The dashed line has
slope -2. The inset shows the values for large-gap SWNTs as a
function of $R_0/\Delta_g^2$. }  \label{fig:pollong}
\end{figure}

All calculations are performed using Quantum-ESPRESSO \cite{PWSCF}
with the PBE approximation and ultra-soft 
pseudo-potentials in a plane-wave basis. A tetragonal unit cell is
set up with periodic-boundary conditions in all three dimensions. A
$k$-point sampling grid of at least 30x1x1 points is used; this is
sufficient to converge polarizabilities to within 3 significant
digits. Atomic configurations are generated using an interatomic
distance of 1.42\AA, obtained from careful relaxation studies
\cite{mounet}. Longitudinal and transverse polarizabilities are
calculated using density-functional perturbation theory (DFPT)
\cite{dfpt} and finite-field or electric-enthalpy approaches
\cite{paolo}, also implemented in our Quantum-ESPRESSO code. Since
we use periodic-boundary conditions, we effectively simulate a
three-dimensional bulk material consisting of a square array of
infinite parallel nanotubes. The longitudinal dielectric response of
an isolated nanotube is characterized by polarizability per unit
length $\alpha_{\parallel}$, which is related to the
separation-dependent bulk dielectric constant $\epsilon_{\parallel}$
using the relation
\begin{equation} \label{eq:eps0}
\epsilon_{\parallel} = 1 + \frac{{4\pi }}{\Omega }\alpha_{\parallel}
\end{equation}
where $\Omega=L^2$ is the cross-sectional area of the unit cell.
From linear-response theory \cite{penn} we expect the static
dielectric constant to depend on the gap as $\epsilon(q) \approx 1 +
\left({{\hbar \omega _p }}/{{\Delta _{g} }} \right)^2$
which suggests via (\ref{eq:eps0}) that $\alpha_{\parallel}\sim
1/\Delta_g^2$. Our calculations confirm this behavior in zigzag
nanotubes, as shown in Fig. \ref{fig:pollong}. As expected, (9,0), (12,0) and
(15,0) nanotubes have the smallest gaps and the largest
$\alpha_{\parallel}$; the inverse-square dependence on the gap
roughly holds over two orders of magnitude. Only the narrowest
nanotubes (7,0) and (8,0) deviate from this trend. The agreement is
particularly accurate for large-gap zigzag nanotubes (3n+1,0) and
(3n+2,0) with $n > 2$. We note in passing that for these SWNTs our
first-principles results can be fitted well with these relations:
$\Delta_g \approx 3.3 / R_0 + 0.06$ and $\alpha_{\parallel} \approx
8.2 R_0^2 + 20.5 $, with $\Delta_g$ in eV and $R_0$ in \AA. Previous
tight-binding studies \cite{benedict} reported values of
$\alpha_{\parallel}$ comparable to ours, and noted a relation
$\alpha_{\parallel}\sim R_0/\Delta_g^2$ which we also observe for
large-gap nanotubes (see Fig. \ref{fig:pollong}). For infinitely-long armchair SWNTs the
longitudinal polarizability per unit length $\alpha_{\parallel}$
diverges since there is no gap in the band dispersions. To get a
sense of scaling we can approximate such nanotubes as metallic
ellipsoids of length $l$ and transverse radius $R$ ($l\gg R$); the
classical result is $\alpha_{\parallel} \sim l^2/[24(\ln(l/R)-1)]$.
For MWNTs, the longitudinal picture remains simple: depolarization
effects along the axis are negligible, and constituent tubes have
very weak dielectric interactions. The total polarizability
$\alpha_{\parallel}^{tot}$ should then simply be the sum of the
polarizabilities of constituent SWNTs; this conclusion is confirmed
by our results in Table \ref{tab:results}.
\begin{table} \caption{\label{tab:results} Radius, band gap, longitudinal and transverse
polarizabilities (per unit length) of carbon nanotubes as a function
of the chiral vector $(n,m)$.}
\begin{ruledtabular}
\begin{tabular}{c|cc|cc}
$(n,m)$ & $R_0$ (\AA)\footnotemark[1]  & $\Delta_g$ (eV) &
$\alpha_{\perp}$ (\AA$^2$)& $\alpha_{\parallel}$ (\AA$^2$)\\
\hline
(7,0) & 2.74 & 0.48&  6.47  & 83.0\\
(8,0) & 3.15 & 0.57 & 7.80 & 104\\
(9,0) & 3.58 & 0.17 & 9.32 & 1460 \\
(10,0) & 3.95 & 0.91 & 10.9  & 142\\
(11,0) & 4.34 & 0.77 & 12.7  & 186\\
(12,0) & 4.73 & 0.087 & 14.3  & 6140\\
(13,0) & 5.09 & 0.72 &16.3  & 224\\
(14,0) & 5.48 & 0.63 & 18.4 & 279\\
(15,0) & 5.88 & 0.041 & 20.3 & 11100\\
(16,0) & 6.27 & 0.61 &22.9 & 326\\
(17,0) & 6.66 & 0.53 & 25.2  & 395\\
(8,0)+(17,0) & - & - &  25.8 & 499\\
(8,0)+(16,0) & - &  - & 23.6 & 427 \\
\hline (4,4) & 2.71 & (0) & 6.41 & $(\infty)$\\
(5,5) & 3.40 &- & 8.71 & -\\
(6,6) & 4.10 & - & 11.6 & -\\
(7,7) & 4.76 & - & 14.7 & - \\
(8,8) & 5.45 & - & 18.1 & - \\
(9,9) &6.12 & - & 21.8 & -\\
(10,10) & 6.78 & - & 26.1 & - \\
(12,12) & 8.14& - & 35.8 & -\\
(14,14) & 9.50 & - & 47.2 & -\\
\end{tabular}
\end{ruledtabular}
\footnotetext[1]{$R_0$ is the radius of the carbon backbone}
\end{table}
\begin{table} \caption{\label{tab:mwnt}Transverse polarizabilities of MWNTs.}
\begin{ruledtabular}
\begin{tabular}{c|cc}
 MWNT & $\alpha_\perp$ (\AA$^2$)(ab-initio) &
 $\alpha_{\perp}$(\AA$^2$) (model)\\
\hline
(8,0)+(17,0) & 25.8 & 25.7 \\
(5,5)+(10,10) & 26.8 & 26.6\\
(4,4)+(12,12) & 36.1 & 36.0 \\
(9,9)+(14,14) & 49.0 & 48.2 \\
(4,4)+(9,9)+(14,14) & 49.1 & 48.3
\end{tabular}
\end{ruledtabular}
\end{table}
\begin{figure}[htb]
\centering \includegraphics[width=3.0 in]{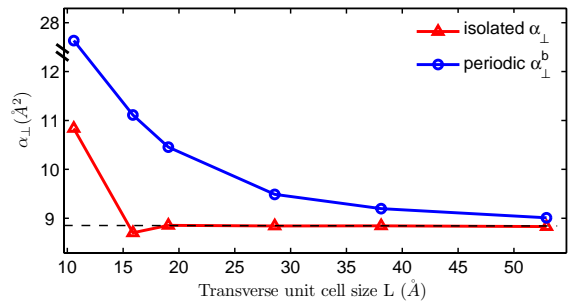} \caption
{Convergence of $\alpha_\perp$ and $\alpha_\perp^b$ with respect to
L for a (5,5) SWNT. The point at L=10.6\AA \ corresponds to a
typical tube-tube separation in a bundle.} \label{fig:conv}
\end{figure}

\begin{figure}[htb]
\centering \includegraphics[width=3.0in]{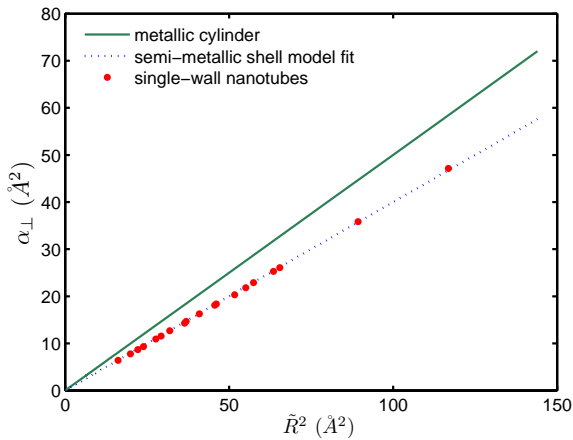} \caption
{Transverse polarizabilities $\alpha_{\perp}$ of armchair and zigzag
nanotubes as a function of $\tilde{R}^2$. The dashed line is the
best-fit result of our semi-metallic shell model; the solid line
$\alpha_\perp=\frac{1}{2}\tilde{R}^2$ corresponds to an ideal
metallic cylinder. } \label{fig:poltr}
\end{figure}

We address the characterization of the transverse dielectric response
in two different ways. First, we calculate with DFPT the dielectric
constant $\epsilon_\perp$, from which the transverse polarizability
$\alpha_\perp$ is extracted. To study non-linear finite-field effects,
we also obtain $\alpha_\perp$ by applying an electric field $E_{out}$
via a sawtooth potential, and computing the total induced dipole
moment per unit length $p_\perp$. In the linear regime the two
approaches are equivalent, and we find an agreement between the two
methods within 1\%. Both DFT-based calculations take into account the
local-field effects, that would be absent in tight-binding
calculations. Again, since calculations provide us with the transverse
response of a periodically-repeated array of nanotubes, it is
necessary to remove the depolarization fields stemming from the
periodic images. In principle one could use
\begin{equation}\label{eq:direct}
 \alpha_\perp^b=\frac{\Omega}{4\pi}(\epsilon_\perp-1)=\frac{p_\perp}{E_{out}}
\end{equation}
for the first and second methods respectively, while taking the
limit $L\to\infty$ for which the depolarization fields vanish. In
practice, these persist for very large inter-tube separations due to
the long range of electrostatic interactions between image tubes.
Computation time grows as $L^2$ at a fixed energy cutoff, quickly
becoming unmanageable without even reaching a converged result; Fig.
\ref{fig:conv} illustrates the slowness of this convergence. It is
clear, however, that at large separations only electrostatic effects
are important, so we can solve this problem using a classical 2D
Clausius-Mossotti correction \cite{CM} relating the single-tube
polarizability $\alpha_\perp$ to the periodic bulk $L$-dependent
value $\alpha_\perp^b$. The relevant conversions are
\begin{equation}\label{eq:CM}
\alpha_{\perp} = \frac{\Omega
}{2\pi}\frac{\epsilon_{\perp}-1}{\epsilon_{\perp}+1}
=\frac{\alpha_\perp^b}{1+\frac{2\pi}{\Omega}\alpha_\perp^b}\!.
\end{equation}
The values of $\alpha_\perp$ obtained from (\ref{eq:CM}) are listed
in Table \ref{tab:results} and plotted in Fig. \ref{fig:poltr} as a
function of the square of the effective outer radius
$\tilde{R}=R_0+1.3$\AA \ (see later discussion). Remarkably,
transverse polarizabilities of both metallic and semiconducting
SWNTs lie on the same curve, which can be fitted by a line
$\alpha_\perp=c\tilde{R}^2$ with slope $c=0.40$. Thus chirality and
longitudinal band structure have a negligible effect on the
transverse dielectric response; this was observed in earlier
calculations \cite{benedict,scuseria,nonlinear} and justified with
symmetry arguments in the single-particle approximation
\cite{benedict}. Recent tight-binding calculations
\cite{levitov,rotkin} predict a small and systematic difference
between polarizabilities of metallic and semiconducting SWNTs;
however, we do not detect these differences in our DFT calculations.

Periodic boundary conditions allow us to easily examine the bulk
dielectric response of nanotube bundles. We compute $\epsilon_\perp$
and $\epsilon_\parallel$ of triangular and square arrays with
inter-tube separation of $d$=3.4\AA \cite{bundles}. The values of
$\epsilon_\parallel$ accurately match those computed from
$\alpha_\parallel$ of isolated nanotubes using (\ref{eq:eps0}), thus
reflecting the additive property of the longitudinal response. In
contrast, transverse response of bundles depends strongly on $d$.
Fig. \ref{fig:conv} illustrates both the benefits of using the
Clausius-Mossotti relation (\ref{eq:CM}) for large $d$, and the
limitation of its applicability when $d$ is small. Whereas the
longitudinal response remains simple, the transverse dielectric
tensor at small $d$ may have sizeable anisotropic and off-diagonal
contributions depending on the combined point-group symmetry of the
nanotube and the lattice. These contributions vanish quickly with
$d$ and do not affect our isolated tube calculations.

\begin{figure}[htb]
\centering \includegraphics[width=3.0in]{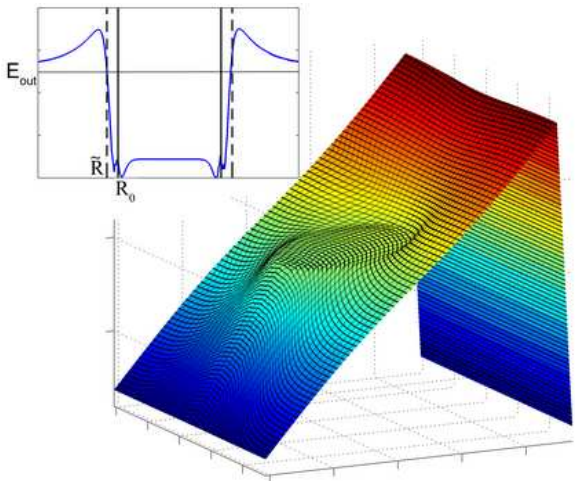}\\ \caption
{Electrostatic potential for a (10,10) SWNT in an applied
homogeneous transverse field $E_{out}$.  The electric field through
the center slice is shown in the inset.} \label{fig:vh}
\end{figure}

By applying a finite transverse field $E_{out}$ we can also study
screening inside a nanotube; we find the inner field $E_{in}$ to be
very uniform, as shown in Fig. \ref{fig:vh}. Another remarkable
feature is that the screening factor $E_{out}/E_{in}\approx 4.4 \pm
0.1$ turns out to be independent of radius and chirality for all
SWNTs. To make physical sense of these general results we look for a
simple electrostatic model that would capture these traits. A solid
dielectric cylinder of radius $\tilde{R}$ and bulk dielectric
constant $\epsilon$ would have polarizability $\alpha_\perp =
\frac{1}{2}\frac{\epsilon-1}{\epsilon+1}\tilde{R}^2$, a uniform
inner field and a screening factor $E_{out}/E_{in}=(\epsilon+1)/2$
independent of radius. This picture, however, does not correspond to
a nanotube, where screening is accomplished by a thin layer of
delocalized $\pi$-electrons. One could then treat a nanotube as a
dielectric cylindrical shell of finite thickness. In this case the
inner field remains uniform, but the screening factor decreases with
increasing radius. To identify an appropriate model that
incorporates all the observed features, we note that in general a
radius-independent uniform inner field is produced by the surface
charge density $\sigma(\phi)=\sigma_0 \cos(\phi)$, where $\phi$ is
the angle measured from the direction of $E_{out}$. The dipole
moment per unit length in this case is $p_\perp=\pi\sigma_0
\tilde{R}^2=\alpha_\perp E_{out}$ and the polarizability is
$\alpha_\perp = \frac{\pi
\sigma_0}{E_{out}}\tilde{R}^2=c\tilde{R}^2$ with $c\leq 1/2$. In a
metallic cylinder $c$ is $1/2$ and the outer field is completely
screened. A best fit of our ab-initio data for SWNTs to this model
(see Fig. \ref{fig:poltr}) yields the slope $c=0.40$ and effective
radius $\tilde{R}=R_0+1.3$\AA \ larger than the radius of the carbon
backbone $R_0$, consistent with the finite thickness of the
electronic charge density distribution. Elementary electrostatic
considerations yield a screening factor
$E_{out}/E_{in}=\frac{1}{1-2c}=5$ in good agreement with our
finite-field calculations and previous estimates
\cite{benedict,levitov}. It should be stressed that the screening
properties of nanotubes, reflected in this model, are {\it neither}
metallic nor insulating. This peculiarity is physically grounded in
the fact that in a single sheet of graphite the screening of Coulomb
interactions is anomalous due to the vanishing density of states at
the Fermi points \cite{gonzalez}. For carbon nanotubes (as opposed
to boron-nitride nanotubes), the semi-metallic nature of
$\pi$-electrons implies that the screening factor is
radius-invariant.

The generalization of this model to the multi-wall case needs to
take into account screening and electrostatic interactions between
layers. Our strategy is to first solve exactly the general problem
of $N$ concentric dielectric cylindrical shells in a uniform field.
We then recover precisely the above single-layer model by treating a
SWNT as a shell of radius $\tilde{R}$, dielectric constant
$\epsilon$ and vanishing thickness $\delta$, and constraining these
parameters by the scale-invariance condition
\begin{equation}\label{eq:scale}
\epsilon \frac{\delta}{\tilde{R}}=\frac{4c}{1-2c}={\rm const}
\end{equation}
that guarantees that the screening factor remains independent of
$\tilde{R}$. Modelling a general MWNT amounts to solving a linear
system of $2N\times2N$ boundary-condition equations \cite{matlab}
containing the best-fit parameters $c$ and $\tilde{R}$ (carried over
from the single-wall case) and subject to constraint
(\ref{eq:scale}). For the double- and triple-wall cases we find
excellent agreement between this model and our ab-initio results
(see Table \ref{tab:mwnt}). We conclude that the present
semi-metallic shell model captures all characteristics of the
transverse dielectric response: uniform inner field,
radius-independent screening factor in SWNTs, and correct
$\alpha_\perp$ for MWNTs. We note also that the largest
contributions to transverse polarizabilities come from the outer few
layers, and inner layers play a negligible role due to a combination
of screening and their smaller radii. Diameter control alone thus
becomes the key growth-parameter determining transverse response.

The finite-field approach is also used to determine the range of
fields for which the transverse dielectric response is linear. The
$(5,5)$ nanotube exhibits precisely linear response with the same
polarizability coefficient to within 3 significant digits for field
magnitudes of 0.05, 0.5, 5 V/nm, the last one being greater than the
experimentally attainable value. This implies that our electrostatic
shell model of transverse response remains valid in the regime of
large applied fields. To study the linearity of longitudinal response,
we minimize directly the electric-enthalpy functional \cite{paolo} to
introduce a finite longitudinal field while preserving
periodic-boundary conditions. We find that the longitudinal response
of the (8,0) nanotube becomes nonlinear by only 5\% at
$E_\parallel$=0.5 V/nm. Nonlinearity is in fact suppressed because
zigzag and armchair (non-chiral) nanotubes are center-symmetric, so
the first hyper-polarizability $\beta$ vanishes by symmetry
\cite{nonlinear}. To estimate the second hyper-polarizability
$\gamma_\parallel$ we compute polarizations at several values of the
field, and fit the result to the expression $P=\alpha_\parallel E +
\gamma_\parallel E^3$. We obtain $\alpha_\parallel=106$ \AA$^2$ (in
agreement with the DFPT result in Table \ref{tab:results}) and
$\gamma_\parallel=3.1\times 10^7$ in atomic units.

We now turn to the question of alignment of nanotubes in a uniform
electric field. The torque on a nanotube of length $l$ at an angle
$\theta$ to the field $\bf{E}$ is
\begin{equation}\label{torque}
\tau  = \left| {{\bf{p}} \times {\bf{E}}} \right| = l \left( {\alpha
_{\parallel}  - \alpha_{\perp}  } \right)E^2 \sin \theta \cos \theta
\end{equation}
The longitudinal and transverse polarizabilities compete with each
other, but our results imply that $\alpha_\parallel>\alpha_\perp$ in
all nanotubes, much more so in metallic and small-gap semiconducting
nanotubes. Indeed, for all nanotubes $\alpha_\perp <
\frac{1}{2}\tilde{R}^2$ whereas for large-gap SWNTs
$\alpha_\parallel \gtrsim 8.2 R_0^2$, and for MWNTs
$\alpha_\parallel$ is additive while $\alpha_\perp$ is not. So
nanotubes of all types will align with the electric field, but by
tuning the value of the field during PECVD growth it may be possible
to selectively grow highly polarizable (e.g. metallic) tubes.

There have also been attempts to separate semiconducting and
metallic nanotubes in solution using inhomogeneous electric fields
\cite{solution}. A polarized nanotube aligned with the field will be
pulled in the direction of or against the field gradient, depending
on its effective dielectric constant $\epsilon_\parallel$ relative
to that of the solvent $\epsilon_s$. Assuming no solvent inside the
nanotube, and approximating it by a solid dielectric cylinder of
radius $R_0$, we obtain from our values of $\alpha_\parallel$ an
effective $\epsilon_\parallel =1+4\alpha_\parallel/R_0^2\approx 30$
for large-gap semiconducting SWNTs and obviously much larger values
for small-gap and metallic tubes. This result is consistent with
findings that only metallic SWNTs are observed deposited on the
electrodes in water ($\epsilon_s=80$), whereas all nanotubes are
drawn towards the electrodes in isopropyl alcohol ($\epsilon_s=18$).

In summary, we studied in detail the dielectric properties of
isolated and bundled SWNTs and MWNTs. In SWNTs, the longitudinal
response is controlled by the band gap, while the transverse
response is sensitive only to the effective radius. In bundles and
MWNTs longitudinal response is additive, while the transverse
response in MWNTs is dominated by the outer few layers. We presented
an accurate scale-invariant electrostatic model of transverse
response, which is intermediate between that of a metal and an
insulator. The authors would like to thank L. S. Levitov for
valuable suggestions. This work was supported by NSF-NIRT
DMR-0304019 and the Singapore-MIT alliance.

\end{document}